\documentclass[apjl,numberedappendix]{emulateapj}

\usepackage[backref,breaklinks,colorlinks,citecolor=blue]{hyperref} 
\usepackage[all]{hypcap} 

\usepackage{graphicx}

\usepackage{natbib}
\usepackage{amsmath}

\newcommand{\ci}{[\ion{C}{1}]}

\newcommand{\cii}{[\ion{C}{2}]}
\newcommand{\mgii}{Mg\ensuremath{\,\textsc{ii}}}
\newcommand{\ciim}{[C\ensuremath{\,\textsc{ii}}]}

\newcommand{\nii}{[\ion{N}{2}]}

\newcommand{\oi}{[\ion{O}{1}]}

\newcommand{\lcii}{L_{\ensuremath{\rm [C\,\textsc{ii}]}}}
\newcommand{\sfrcii}{{\rm SFR}_{\ensuremath{\rm [C\,\textsc{ii}]}}}

\newcommand{\sfrtir}{{\rm SFR}_{\ensuremath{\rm TIR}}}

\newcommand{\lfir}{L_{\ensuremath{\rm FIR}}}
\newcommand{\ltir}{L_{\ensuremath{\rm TIR}}}

\newcommand{\water}{H$_2$O}

\def\kms{{\rm\,km\,s^{-1}}}
\def\jkms{{\rm\,Jy\,km\,s^{-1}}}
\def\mjybeam{{\rm\,mJy\,beam^{-1}}}

\def\kpc{{\rm\,kpc}}

\def\msyr{M_\odot \, \rm yr^{-1}}
\newcommand{\lsun}{$L_\sun$}

\shorttitle{\cii\ and dust in  a bright quasar at $z=6.54$}
\shortauthors{Ba\~{n}ados et al.}
\begin{document}

\title{Bright  \cii\ 158\,$\mu$\lowercase{m} emission in a quasar host galaxy at $\lowercase{z}=6.54$\altaffilmark{*} }

\author{
E.~Ba\~{n}ados\altaffilmark{1},
R.~Decarli\altaffilmark{1},
F.~Walter\altaffilmark{1},
B.P.~Venemans\altaffilmark{1},
E.P.~Farina\altaffilmark{1},
X.~Fan\altaffilmark{2}
}

\altaffiltext{1}{Max Planck Institut f\"ur Astronomie, K\"onigstuhl 17, D-69117, Heidelberg, Germany}%
\altaffiltext{2}{Steward Observatory, The University of Arizona, 933 North Cherry Avenue, Tucson, AZ 85721--0065, USA}
\email{banados@mpia.de}
\altaffiltext{*}{Based on observations carried out under project number E14AG with the IRAM NOEMA Interferometer. IRAM is supported by INSU/CNRS (France), MPG (Germany) and IGN (Spain).
}

\begin{abstract}
The \cii\ 158$\,\mu$m fine-structure line is known to trace regions of active
 star formation and is the main coolant of the cold, neutral atomic medium. In this \textit{Letter}, we report a strong detection of the \cii\ line 
 in the host galaxy of the brightest quasar known at $z>6.5$, the Pan-STARRS1 selected quasar PSO J036.5078+03.0498 (hereafter P036+03), 
 using the IRAM NOEMA millimeter interferometer. 
 Its  \cii\ and total far-infrared luminosities are $(5.8 \pm 0.7) \times 10^9 \, $\lsun\ and $(7.6\pm1.5) \times 10^{12}\,$\lsun, respectively. 
 This results in a $\lcii /\ltir$ ratio of $\sim 0.8\times 10^{-3}$, which is at the high end for those found for active galaxies, though it is lower
 than the average found in typical main sequence galaxies at $z\sim 0$.
 We also report a tentative additional line which we identify as a blended emission from the $3_{22} - 3_{13}$ and $5_{23} - 4_{32}$ H$_2$O transitions.
 If confirmed, this would be the most distant detection of water emission to date.
 P036+03 rivals the current prototypical luminous J1148+5251 quasar at $z=6.42$, in both rest-frame UV and \cii\ luminosities.
 Given its brightness and because it is visible from both hemispheres (unlike J1148+5251), P036+03 has the potential
 of becoming an important laboratory for the study of star  formation and of the interstellar medium 
 only $\sim 800$\,Myr after the Big Bang.
\end{abstract}


\vfil
\eject
\clearpage

\section{INTRODUCTION}
\label{sec:intro}

The study of star formation in the first galaxies back in the epoch of reionization ($z>6.5$) is one of the main challenges in current 
observational cosmology.

Several groups have tried to identify the \cii\ 158\,$\mu$m fine-structure transition line (hereafter \cii) in the first
spectroscopically identified galaxies at these redshifts. The \cii\ line is the dominant coolant 
of interstellar neutral gas and is one of the brightest lines in the spectrum of star-forming galaxies, accounting for up to $\sim 1\%$
of the total far-infrared (FIR) luminosity in $z\sim 0$ main sequence galaxies. Studies of the \cii\ line may also provide key insights into galaxy kinematics 
at the highest redshifts \citep[see][for a review]{carilli2013}. However, thus far, there are no convincing detections of \cii\ emission 
from star-forming galaxies during the epoch of reionization
(\citealt{walter2012b,ouchi2013,kanekar2013,gonzalez-lopez2014,ota2014,schaerer2015,watson2015};
but see also the recent work of  \citealt{maiolino2015}).

On the other hand, the \cii\ line has been identified in a number of more extreme objects such as submillimeter galaxies and quasar host galaxies
at $z\sim 6$ \cite[][]{maiolino2005, walter2009,riechers2013,wang2013, willott2013b, willott2015}. These objects are most likely the progenitors of massive early-type galaxies
seen in the present universe. Thus, luminous quasars are thought to be important tools to pinpoint the locations of these extreme galaxies in the early
universe. Indeed, the  \cii\ line has been detected even in the host galaxy of the currently most distant quasar
known at $z=7.1$ \citep{venemans2012}. 

For more than a decade, the quasar SDSS J1148+5251 at $z=6.42$ (hereafter J1148+5251), discovered by \cite{fan2003},
has been the brightest quasar known at $z>6$ ($M_{1450}=-27.8$). 
As such, it is by far the best studied quasar at these redshifts, being fundamental in shaping our current 
understanding of the universe when it was less than 1\,Gyr old.
It was the first source detected in \cii\ at $z>0.1$ \citep{maiolino2005}, and for several years it remained the only $z>6$ quasar detected also
in CO emission \citep{walter2003, walter2004,bertoldi2003b,riechers2009}.
The \cii\ line in this quasar has been studied in detail \citep[e.g.,][]{walter2009, maiolino2012}.
Neutral carbon \ci\ has also been detected \citep{riechers2009}, while deep limits on the \nii\ line 
 could also be placed \citep{walter2009b}.

Recently, and after more than 10 years, three quasars that rival (and in one case even surpass) J1148+5251 in UV luminosity have been discovered.
These quasars are SDSS J0100+2802 \citep[$z=6.30; M_{1450}=-29.3$;][]{wu2015}, ATLAS J025.6821--33.4627 \citep[$z=6.31; M_{1450}=-27.8$;][]{carnall2015},
and  PSO J036.5078+03.0498  \citep[hereafter P036+03; $z=6.527; M_{1450}=-27.4$;][]{venemans2015}.

In this \textit{Letter}, we focus on P036+03, which is the brightest quasar known at $z>6.5$ (see Figure 3 in \citealt{venemans2015}).
We report a bright 
detection of the \cii\ line with the IRAM NOEMA interferometer.
We also detect the underlying dust far-infrared continuum emission and present a tentative detection of a water transition in this quasar. 
As a result, P036+03 (that is also accessible from telescopes in the southern hemisphere) competes with the current archetypal high-redshift quasar 
J1148+5251 in both rest-frame UV and \cii\ luminosities.

We employ a cosmology with $H_0 = 67.8 \,\mbox{km s}^{-1}$ Mpc$^{-1}$, $\Omega_M = 0.307$, and $\Omega_\Lambda = 0.691$ \citep[][]{planck2014}.
This implies that for $z=6.5412$, the proper spatial scale is $5.6\,\kpc\,$arcsec$^{-1}$ and 
the age of the universe is 833\,Myr, 
i.e., 6\% of its present age.

\section{Observations and results}
\label{sec:observations}

We use the IRAM NOEMA interferometer to observe the \cii\ line in the quasar P036+03, as a director's discretionary time proposal (ID: E14AG). 
The tuning frequency is 252.496\,GHz 
which corresponds to the expected \cii\ frequency ($\nu_{rest}=1900.539\,$GHz) based on the quasar redshift estimated from
the \mgii\ line:
$z_{\rm \mgii}=6.527$ \citep{venemans2015}. 
The observations were carried out in 2015 February 09 and 10 in the D configuration with
a total time on source of only 3.55 hr (5 antenna equivalent).
The synthesized beam size is $2.2\arcsec \times 1.6\arcsec$.
The data were calibrated and analyzed with the IRAM GILDAS\footnote{\url{http://www.iram.fr/IRAMFR/GILDAS}} software package. 

\begin{figure}[ht]
\begin{center}
\vspace{-20pt}
\includegraphics[scale=0.4]{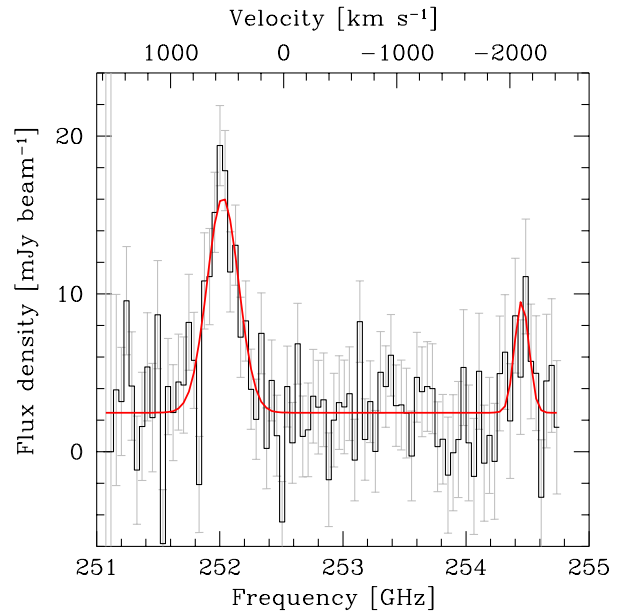}
\vspace{-60pt}
 \caption{NOEMA spectrum at 1\,mm (observed) wavelengths of P036+03. The tuning frequency was centered on the expected position of \cii\ based on the 
 redshift determined from the \mgii\ line ($\nu_{obs}=252.496$\,GHz). This tuning frequency is shown here as zero point of the velocity scale.
 The red solid curve is a fit to the data consisting of two Gaussians plus continuum. The \cii\ line is detected at high significance and is 
 redshifted by $566 \pm 72 \kms$  with respect to the \mgii\ redshift ($z_{\rm \ciim}=6.54122$, $z_{\rm \mgii}=6.527$). A second, fainter line is observed at $254.45\pm0.14\,$GHz.  We find that the 
 most likely identification for this line is a blended emission from  the adjacent $3_{22} - 3_{13}$ and $5_{23} - 4_{32}$ H$_2$O emission lines (see text).
 \label{fig:ciispc}}
 \end{center}
 \end{figure}

\begin{figure*}[ht]
\begin{center}
\includegraphics[scale=0.4]{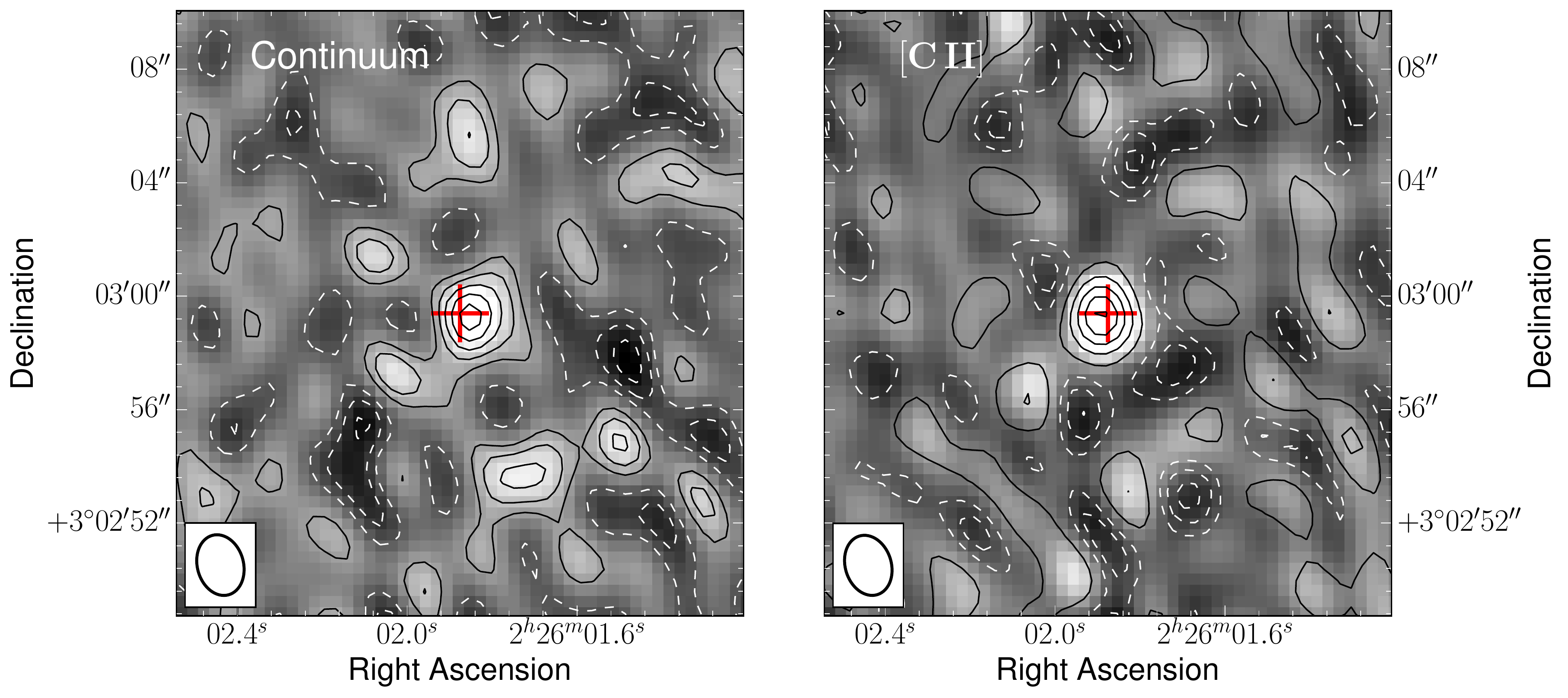}
 \caption{NOEMA cleaned maps of the $z=6.5412$ quasar P036+03. The red cross in each panel shows the near-infrared location of the quasar.
 The beam size of $2.2\arcsec \times 1.6\arcsec$ is shown at the bottom left of each panel. 
 \textit{Left:} Far-infrared continuum emission obtained from line--free channels in the frequency range: 254.274\,GHz -- 252.524\,GHz (i.e., $2100\kms$, see Figure \ref{fig:ciispc}).
 The contours correspond to $-3\sigma$, $-2\sigma$, $-1\sigma$ (white dashed lines), $1\sigma$, $2\sigma$, $3\sigma$, $4\sigma$, and $5\sigma$ 
 (black solid lines), with $\sigma$ being the rms noise of $0.418\mjybeam$.
 \textit{Right:} Continuum subtracted \cii\ emission integrated from the channels 251.372\,GHz to 251.782\,GHz (i.e., $750\kms$).
 The contours are $-3\sigma$, $-2\sigma$, $-1\sigma$ (white dashed lines), $1\sigma$, $3\sigma$, $5\sigma$, $7\sigma$, and $9\sigma$ 
 (black solid lines), with $\sigma=0.85\mjybeam$. The excess of $3\sigma$ sources in the maps is due to the modest 
 $u,v$ coverage in our data. 
 \label{fig:maps}}
 \end{center}
 \end{figure*}

The final NOEMA spectrum of P036+03 is shown in Figure~\ref{fig:ciispc}.
 Two lines are visible, a bright one, and a second fainter line on top of the underlying dust
continuum.
We fit two Gaussians plus continuum to the data and identify \cii\ as the brightest of these lines.
The \cii\ line peaks at $252.02 \pm 0.06\,$GHz, yielding a redshift of $z_{\rm \ciim}=6.54122 \pm 0.0018$ 
(i.e., redshifted by  $566 \pm 72 \kms$  with respect to the \mgii\ redshift). The line has a FWHM$=360 \pm 50\kms$ and a velocity integrated flux of $5.2 \pm 0.6\jkms$. This corresponds to a \cii\ luminosity of
$\lcii = (5.8\pm0.7) \times 10^9 $\,\lsun which is a factor of $\sim$1.4 brighter than the \cii\ line in J1148+5251
\citep[$4.2 \pm 0.4 \times 10^9\,$\lsun;][]{maiolino2005,walter2009}. The second line peaks at $254.45 \pm 0.14\,$GHz, i.e., blueshifted by 
$-2863 \pm 178 \kms$ with respect to the \cii\ line.  This fainter line has a FWHM$=160\pm100\, \kms$
and an integrated flux of $1.2\pm 0.7 \jkms$. At our resolution, the most likely identification for this line is a 
blended water vapor emission from the adjacent $3_{22} - 3_{13}$ and $5_{23} - 4_{32}$ H$_2$O transitions ($\nu_{rest} = 1919.359\,$GHz and 
$\nu_{rest} = 1918.485\,$GHz, respectively).  
This tentative blended water emission has a velocity offset of $576 \pm 165 \,\kms$ with respect to its predicted value by the \ion{Mg}{2} redshift.
 This offset is in good agreement with the velocity offset found for the \cii\ line, reinforcing the identification of the water emission.
Figure \ref{fig:maps} shows the maps of the far-infrared continuum emission and the continuum subtracted \cii\ emission.

Fitting the continuum flux density yields a value of $2.5 \pm 0.5 \,$mJy. We convert the measured dust continuum flux density to 
far-infrared luminosity, $\lfir$, by modeling the FIR emission with an optically thin gray body spectrum and scale the model 
to match the observed dust continuum flux density. We use typical parameters in the model to reproduce the SED for high-redshift
quasar host galaxies, i.e., a dust temperature of $T_d = 47 \,$K and a emissivity index of $\beta=1.6$ 
 \citep[following for example,][]{venemans2012, willott2015}.
 Integrating the model from rest-frame $42.5\, \mu$m to $122.5\, \mu$m we obtain the following $\lfir$ for P036+03: $\lfir = (5.4 \pm 1.1) \times 10^{12}\,$\lsun.
 Integrating the model from rest-frame  $8\,\mu $m to $1000\,\mu$m we get a total infrared luminosity,
 $\ltir$, of $\ltir=(7.6 \pm 1.5) \times 10^{12} \,$\lsun.
 We can use the total infrared luminosity to estimate the star formation rate using the relation 
 $\sfrtir (\msyr) = \delta_{MF} \times 1.0 \times 10^{-10}$\,$\ltir$ (\lsun) \citep[][and references therein]{carilli2013}, where $\delta_{MF}$ depends on the stellar population.
 We use $\delta_{MF}=1.0$ which is appropriate for a Chabrier IMF. For P036+03, this relation results in $\sfrtir = 760 \pm 150\, \msyr$.
Alternatively, \cite{sargsyan2014} show that even in sources dominated by AGN, the star formation rate can be estimated from $\lcii$ to
a precision of 
$\sim 50\%$ with the relation $\sfrcii (\msyr) = 10^{-7}$\,$\lcii$ (\lsun). This relation yields $\sfrcii = 580 \pm 70\, \msyr$, 
broadly consistent with the estimate above.

 \vspace{-0.5cm}
\section{Discussion}
 
\subsection{\cii\ - IR luminosity relation}
\begin{figure}[ht]
\begin{center}
\includegraphics[scale=0.59]{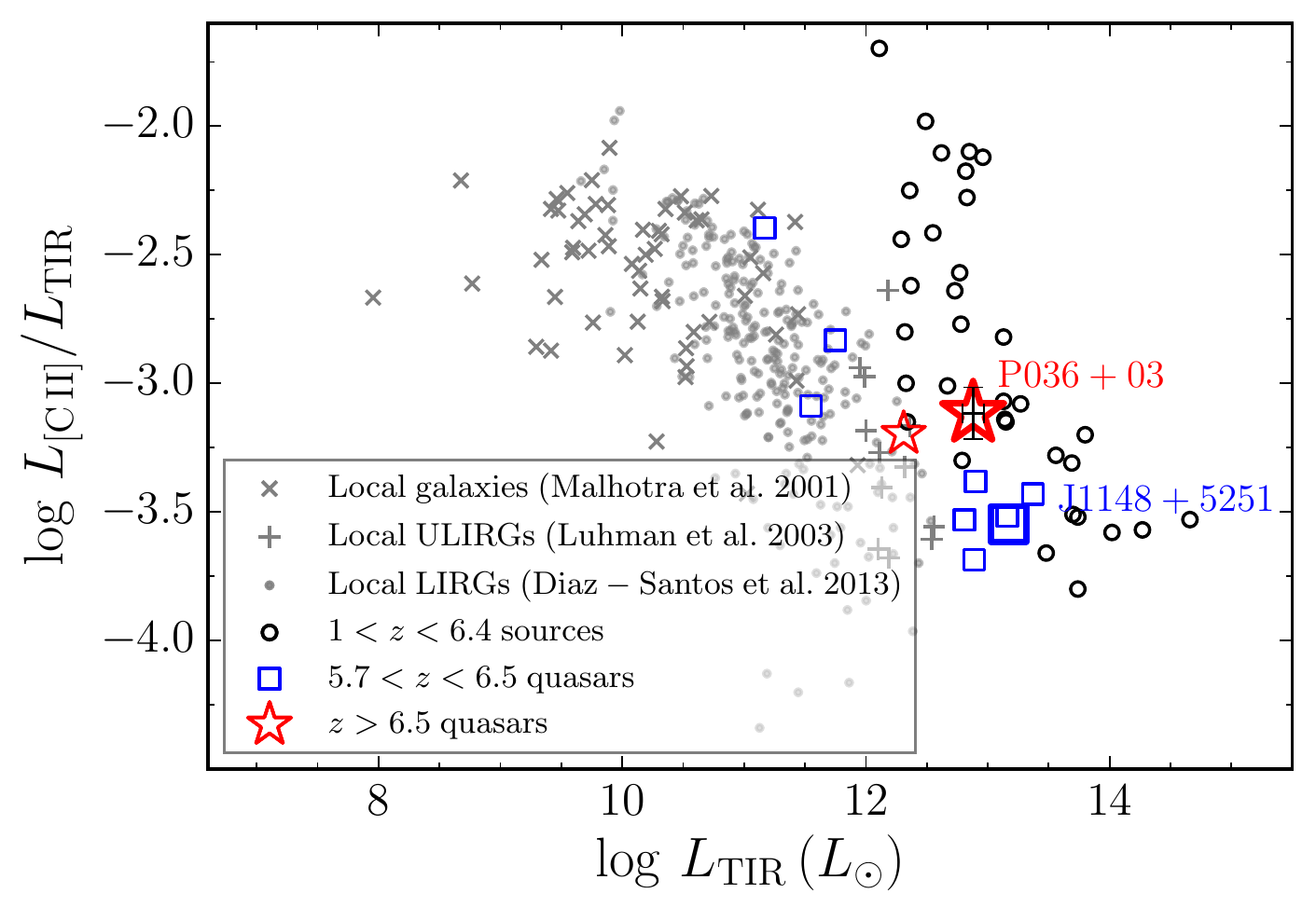}
 \caption{The ratio of \cii\ to total IR luminosity (TIR; $8\,\mu $m to $1000\,\mu$m) vs TIR luminosity. 
 The sources at $1.0<z<6.4$ include the compilation of star-forming galaxies, submillimiter galaxies, and quasars by \cite{delooze2014} and
 8 star-forming galaxies recently reported by \cite{brisbin2015}.
 The $5.7 <z<6.5$ quasars are from \cite{bertoldi2003,maiolino2005,walter2009, wang2013, willott2013b, willott2015}.
 The $z>6.5$ quasars are P036+03 (this work) and J1120+0641 at $z=7.1$ \citep{venemans2012}. 
 Luminosities have been calculated with the cosmology used in this paper. The TIR luminosities for the $z>5.7$ quasars have been 
 consistently calculated as for P036+03 (see text Section \ref{sec:observations}).
 The FIR ($42.5\,\mu $m to $122.5\,\mu$m) luminosities from \cite{diaz-santos2013} have been converted to TIR luminosities using a common
 conversion factor of 1.75 \citep{calzetti2000}.
 Error bars are only plotted for P036+03 to enhance
 the clarity of the figure.
  \label{fig:ciifir}}
 \end{center}
 \end{figure}

It has been shown that in local star-forming galaxies the ratio between the \cii\ and IR luminosities is
about 0.1\% -- 1\% but it decreases in galaxies
with larger IR luminosities ($\ltir > 10^{11}$\,\lsun) such as LIRGs and ULIRGs. This has been known as the ``\cii\ deficit''.
Similar deficits have been observed in other far-infrared lines such as \oi\,63.2\,$\mu$m, \oi\,145\,$\mu$m, and \nii\,122\,$\mu$m.
As a consequence, these trends are now being referred to as the ``far-infrared lines deficit'' \citep{gracia-carpio2011}.
In Figure \ref{fig:ciifir} we show the ratio of \cii\ to TIR luminosity for different type of sources. 
The ``\cii\ deficit'' can be clearly appreciated in the local samples: star-forming galaxies \citep{malhotra2001},  
LIRGs \citep{diaz-santos2013}, and  ULIRGs \citep{luhman2003}. 
The sample of IR-luminous star-forming galaxies, submillimeter galaxies, and quasars at $z>1$ compiled by \cite{delooze2014} and \cite{brisbin2015}
shows a larger scatter in the \cii\ to TIR luminosity ratio ($ \sim 10^{-4} - 10^{-1.5}$). 
The sample of $z\sim 6$ quasars with high IR luminosities studied by \cite{maiolino2005,walter2009b,wang2013} shows small \cii\ to FIR luminosity
ratios, similar to what is observed in local ULIRGs ($ \sim 10^{-3.5} $). One interpretation of such low ratios could be in part due to 
the central AGN contributing to the IR luminosity \citep{wang2013}, but note that there is a big literature dealing with possible explanations
for the deficit \citep[e.g.,][]{malhotra2001, gracia-carpio2011,diaz-santos2013}. 
\cite{willott2015} studied a sample of $z\sim 6$ quasars two order of magnitudes fainter in the infrared ($\ltir= 10^{11 - 12}\,$\lsun).
These quasars show ratios consistent with the local star-forming galaxies ($\sim 10^{-3} - 10^{-2.5}$). \cite{venemans2012} report 
that the quasar J1120+0641 at $z=7.1$ has a
TIR luminosity that is between the \cite{willott2015} sample and the IR-bright $z\sim  6$ quasars. The $\lcii / {\ltir}$ ratio for J1120+0641 is more consistent
with local star-forming galaxies than to ULIRGs.
The quasar of the present study, P036+03, has a $\ltir$ comparable to other $z\sim 6$ quasars ($\ltir \sim 10^{13} \,$\lsun). However, it has a significant larger
$\lcii / {\ltir}$ ratio ($\sim 0.8 \times 10^{-3}$), similar to J1120+0641, and consistent with the lower-end of local star-forming galaxies. This ratio is 2.7 times larger than the ratio reported for J1148+5251.

Together, this emphasizes the differences between the host galaxies of the most distant quasars. 
However, the number of quasars with \cii\ and dust far-infrared continuum information is still to small to derive statistically meaningful trends or correlations.

 \subsection{Water vapor detection}
 The water vapor lines arise in the warm, dense phase of the ISM. 
They are one of the main coolants of this phase, thus playing a key role in
 the fragmentation and collapse of gas clouds \citep{carilli2013}.  To date, there are only a handful of water vapor lines
 reported at high redshift, mostly in strongly magnified sources \citep[e.g.,][]{vanderwerf2011,omont2013b}
 and in one extreme submillimeter galaxy at $z=6.3$ \citep{riechers2013}, being the most distant detection of interstellar water to date.
 Recently, \cite{bialy2015} show that high abundances of water vapor could exist in galaxies
 with extremely low metallicity ($\sim 10^{-3}$ solar), as expected for the first generations of galaxies at high redshifts.
 
As discussed in Section \ref{sec:observations}, we identify a tentative detection of 
blended water vapor emission from the $3_{22} - 3_{13}$ and $5_{23} - 4_{32}$ H$_2$O transitions (see Figure \ref{fig:ciispc}).
Owing to the low significance of the detection is difficult to make any strong conclusions, especially because the water emission is
complex and hard to interpret if only one transition is measured \citep{gonzalez-alfonso2014}.
However, additional water transitions could be detectable in this source. For example, the 
$3_{21} - 3_{12}$ \water\ transition ($\nu_{rest}=1162.912\,$GHz) at $z=6.5412$ is shifted into the 2\,mm band and 
could simultaneously be observed with the CO(J$=10\to9$) line ($\nu_{rest}=1151.985\,$GHz).
The detection of these transitions would lead to an improved excitation model for one of the earliest
galaxies known in the universe and to the confirmation of the most distant water detection to date.

\section{Summary}
\label{sec:summary}
We detect a bright \cii\ emission and the underlying continuum emission in P036+03, 
the most luminous quasar known so far at $z>6.5$. A second line is tentatively detected which we identify
with blended water emission from the adjacent $3_{22} - 3_{13}$ and $5_{23} - 4_{32}$ H$_2$O transitions.
If confirmed, this would be the highest-redshift detection of water vapor available to date.

The \cii\ to TIR luminosity ratio in P036+03 is $\sim 0.8 \times 10^{-3}$, i.e., consistent with low-$L_{TIR}$ star forming galaxies and  
high compared to other IR-luminous quasars at $z>6$. This ratio is a factor $\sim$ 3 higher than in J1148+5251, and consistent with the ratio in the most
distant quasar known to date (Figure \ref{fig:ciifir}), emphasizing the diversity of quasar host galaxies in the early universe.

P036+03 rivals the current archetypal high-redshift quasar, J1148+5251 ($z=6.42$), in both UV and
\cii\ luminosities. Thanks to its brightness and given its convenient equatorial location, P036+03 can be studied from facilities
 in both hemispheres (in contrast to the northern source 
J1148+5251). For example, P036+03 would be an ideal target for high-resolution ALMA imaging to study the ISM physics and to search for signs of mergers,
outflows, and feedback in one of the earliest galaxies in the universe. 
This quasar will thus likely become one of the key targets to study star formation and the ISM in the early universe.

\acknowledgments

We are grateful to K. Schuster for the generous allocation of discretionary time. 
We thank J.M. Winters for his support in the implementation of this program.
E.B. thanks C. Ferkinhoff for useful discussions.

E.B. is a member of the IMPRS for Astronomy \& Cosmic Physics at the University of Heidelberg, Germany.

R.D. is supported by the DFG priority program 1573 `The physics of the interstellar medium'.

B.P.V. and E.P.F acknowledge funding through the ERC grant `Cosmic Dawn'.

{\it Facility:} \facility{IRAM:Interferometer}.


\begin{thebibliography}{}
\expandafter\ifx\csname natexlab\endcsname\relax\def\natexlab#1{#1}\fi

\bibitem[{{Bertoldi} {et~al.}(2003{\natexlab{a}}){Bertoldi}, {Carilli}, {Cox},
  {Fan}, {Strauss}, {Beelen}, {Omont}, \& {Zylka}}]{bertoldi2003}
{Bertoldi}, F., {Carilli}, C.~L., {Cox}, P., {et~al.} 2003{\natexlab{a}}, \aap,
  406, L55

\bibitem[{{Bertoldi} {et~al.}(2003{\natexlab{b}}){Bertoldi}, {Cox}, {Neri},
  {Carilli}, {Walter}, {Omont}, {Beelen}, {Henkel}, {Fan}, {Strauss}, \&
  {Menten}}]{bertoldi2003b}
{Bertoldi}, F., {Cox}, P., {Neri}, R., {et~al.} 2003{\natexlab{b}}, \aap, 409,
  L47

\bibitem[{{Bialy} {et~al.}(2015){Bialy}, {Sternberg}, \& {Loeb}}]{bialy2015}
{Bialy}, S., {Sternberg}, A., \& {Loeb}, A. 2015, ArXiv e-prints,
  arXiv:1503.03475

\bibitem[{{Brisbin} {et~al.}(2015){Brisbin}, {Ferkinhoff}, {Nikola},
  {Parshley}, {Stacey}, {Spoon}, {Hailey-Dunsheath}, \& {Verma}}]{brisbin2015}
{Brisbin}, D., {Ferkinhoff}, C., {Nikola}, T., {et~al.} 2015, \apj, 799, 13

\bibitem[{{Calzetti} {et~al.}(2000){Calzetti}, {Armus}, {Bohlin}, {Kinney},
  {Koornneef}, \& {Storchi-Bergmann}}]{calzetti2000}
{Calzetti}, D., {Armus}, L., {Bohlin}, R.~C., {et~al.} 2000, \apj, 533, 682

\bibitem[{{Carilli} \& {Walter}(2013)}]{carilli2013}
{Carilli}, C.~L., \& {Walter}, F. 2013, \araa, 51, 105

\bibitem[{{Carnall} {et~al.}(2015){Carnall}, {Shanks}, {Chehade}, {Fumagalli},
  {Rauch}, {Irwin}, {Gonzalez-Solares}, {Findlay}, \& {Metcalfe}}]{carnall2015}
{Carnall}, A.~C., {Shanks}, T., {Chehade}, B., {et~al.} 2015, ArXiv e-prints,
  arXiv:1502.07748

\bibitem[{{De Looze} {et~al.}(2014){De Looze}, {Cormier}, {Lebouteiller},
  {Madden}, {Baes}, {Bendo}, {Boquien}, {Boselli}, {Clements}, {Cortese},
  {Cooray}, {Galametz}, {Galliano}, {Graci{\'a}-Carpio}, {Isaak}, {Karczewski},
  {Parkin}, {Pellegrini}, {R{\'e}my-Ruyer}, {Spinoglio}, {Smith}, \&
  {Sturm}}]{delooze2014}
{De Looze}, I., {Cormier}, D., {Lebouteiller}, V., {et~al.} 2014, \aap, 568,
  A62

\bibitem[{{D{\'{\i}}az-Santos} {et~al.}(2013){D{\'{\i}}az-Santos}, {Armus},
  {Charmandaris}, {Stierwalt}, {Murphy}, {Haan}, {Inami}, {Malhotra},
  {Meijerink}, {Stacey}, {Petric}, {Evans}, {Veilleux}, {van der Werf}, {Lord},
  {Lu}, {Howell}, {Appleton}, {Mazzarella}, {Surace}, {Xu}, {Schulz},
  {Sanders}, {Bridge}, {Chan}, {Frayer}, {Iwasawa}, {Melbourne}, \&
  {Sturm}}]{diaz-santos2013}
{D{\'{\i}}az-Santos}, T., {Armus}, L., {Charmandaris}, V., {et~al.} 2013, \apj,
  774, 68

\bibitem[{{Fan} {et~al.}(2003){Fan}, {Strauss}, {Schneider}, {Becker}, {White},
  {Haiman}, {Gregg}, {Pentericci}, {Grebel}, {Narayanan}, {Loh}, {Richards},
  {Gunn}, {Lupton}, {Knapp}, {Ivezi{\'c}}, {Brandt}, {Collinge}, {Hao},
  {Harbeck}, {Prada}, {Schaye}, {Strateva}, {Zakamska}, {Anderson},
  {Brinkmann}, {Bahcall}, {Lamb}, {Okamura}, {Szalay}, \& {York}}]{fan2003}
{Fan}, X., {Strauss}, M.~A., {Schneider}, D.~P., {et~al.} 2003, \aj, 125, 1649

\bibitem[{{Gonz{\'a}lez-Alfonso} {et~al.}(2014){Gonz{\'a}lez-Alfonso},
  {Fischer}, {Aalto}, \& {Falstad}}]{gonzalez-alfonso2014}
{Gonz{\'a}lez-Alfonso}, E., {Fischer}, J., {Aalto}, S., \& {Falstad}, N. 2014,
  \aap, 567, A91

\bibitem[{{Gonz{\'a}lez-L{\'o}pez} {et~al.}(2014){Gonz{\'a}lez-L{\'o}pez},
  {Riechers}, {Decarli}, {Walter}, {Vallini}, {Neri}, {Bertoldi}, {Bolatto},
  {Carilli}, {Cox}, {da Cunha}, {Ferrara}, {Gallerani}, \&
  {Infante}}]{gonzalez-lopez2014}
{Gonz{\'a}lez-L{\'o}pez}, J., {Riechers}, D.~A., {Decarli}, R., {et~al.} 2014,
  \apj, 784, 99

\bibitem[{{Graci{\'a}-Carpio} {et~al.}(2011){Graci{\'a}-Carpio}, {Sturm},
  {Hailey-Dunsheath}, {Fischer}, {Contursi}, {Poglitsch}, {Genzel},
  {Gonz{\'a}lez-Alfonso}, {Sternberg}, {Verma}, {Christopher}, {Davies},
  {Feuchtgruber}, {de Jong}, {Lutz}, \& {Tacconi}}]{gracia-carpio2011}
{Graci{\'a}-Carpio}, J., {Sturm}, E., {Hailey-Dunsheath}, S., {et~al.} 2011,
  \apjl, 728, L7

\bibitem[{{Kanekar} {et~al.}(2013){Kanekar}, {Wagg}, {Ram Chary}, \&
  {Carilli}}]{kanekar2013}
{Kanekar}, N., {Wagg}, J., {Ram Chary}, R., \& {Carilli}, C.~L. 2013, \apjl,
  771, L20

\bibitem[{{Luhman} {et~al.}(2003){Luhman}, {Satyapal}, {Fischer}, {Wolfire},
  {Sturm}, {Dudley}, {Lutz}, \& {Genzel}}]{luhman2003}
{Luhman}, M.~L., {Satyapal}, S., {Fischer}, J., {et~al.} 2003, \apj, 594, 758

\bibitem[{{Maiolino} {et~al.}(2005){Maiolino}, {Cox}, {Caselli}, {Beelen},
  {Bertoldi}, {Carilli}, {Kaufman}, {Menten}, {Nagao}, {Omont}, {Wei{\ss}},
  {Walmsley}, \& {Walter}}]{maiolino2005}
{Maiolino}, R., {Cox}, P., {Caselli}, P., {et~al.} 2005, \aap, 440, L51

\bibitem[{{Maiolino} {et~al.}(2012){Maiolino}, {Gallerani}, {Neri}, {Cicone},
  {Ferrara}, {Genzel}, {Lutz}, {Sturm}, {Tacconi}, {Walter}, {Feruglio},
  {Fiore}, \& {Piconcelli}}]{maiolino2012}
{Maiolino}, R., {Gallerani}, S., {Neri}, R., {et~al.} 2012, \mnras, 425, L66

\bibitem[{{Maiolino} {et~al.}(2015){Maiolino}, {Carniani}, {Fontana},
  {Vallini}, {Pentericci}, {Ferrara}, {Vanzella}, {Grazian}, {Gallerani},
  {Castellano}, {Cristiani}, {Brammer}, {Santini}, {Wagg}, \&
  {Williams}}]{maiolino2015}
{Maiolino}, R., {Carniani}, S., {Fontana}, A., {et~al.} 2015, ArXiv e-prints,
  arXiv:1502.06634

\bibitem[{{Malhotra} {et~al.}(2001){Malhotra}, {Kaufman}, {Hollenbach},
  {Helou}, {Rubin}, {Brauher}, {Dale}, {Lu}, {Lord}, {Stacey}, {Contursi},
  {Hunter}, \& {Dinerstein}}]{malhotra2001}
{Malhotra}, S., {Kaufman}, M.~J., {Hollenbach}, D., {et~al.} 2001, \apj, 561,
  766

\bibitem[{{Omont} {et~al.}(2013){Omont}, {Yang}, {Cox}, {Neri}, {Beelen},
  {Bussmann}, {Gavazzi}, {van der Werf}, {Riechers}, {Downes}, {Krips}, {Dye},
  {Ivison}, {Vieira}, {Wei{\ss}}, {Aguirre}, {Baes}, {Baker}, {Bertoldi},
  {Cooray}, {Dannerbauer}, {De Zotti}, {Eales}, {Fu}, {Gao}, {Gu{\'e}lin},
  {Harris}, {Jarvis}, {Lehnert}, {Leeuw}, {Lupu}, {Menten}, {Micha{\l}owski},
  {Negrello}, {Serjeant}, {Temi}, {Auld}, {Dariush}, {Dunne}, {Fritz},
  {Hopwood}, {Hoyos}, {Ibar}, {Maddox}, {Smith}, {Valiante}, {Bock},
  {Bradford}, {Glenn}, \& {Scott}}]{omont2013b}
{Omont}, A., {Yang}, C., {Cox}, P., {et~al.} 2013, \aap, 551, A115

\bibitem[{{Ota} {et~al.}(2014){Ota}, {Walter}, {Ohta}, {Hatsukade}, {Carilli},
  {da Cunha}, {Gonz{\'a}lez-L{\'o}pez}, {Decarli}, {Hodge}, {Nagai}, {Egami},
  {Jiang}, {Iye}, {Kashikawa}, {Riechers}, {Bertoldi}, {Cox}, {Neri}, \&
  {Weiss}}]{ota2014}
{Ota}, K., {Walter}, F., {Ohta}, K., {et~al.} 2014, \apj, 792, 34

\bibitem[{{Ouchi} {et~al.}(2013){Ouchi}, {Ellis}, {Ono}, {Nakanishi}, {Kohno},
  {Momose}, {Kurono}, {Ashby}, {Shimasaku}, {Willner}, {Fazio}, {Tamura}, \&
  {Iono}}]{ouchi2013}
{Ouchi}, M., {Ellis}, R., {Ono}, Y., {et~al.} 2013, \apj, 778, 102

\bibitem[{{Planck Collaboration} {et~al.}(2014){Planck Collaboration}, {Ade},
  {Aghanim}, {Armitage-Caplan}, {Arnaud}, {Ashdown}, {Atrio-Barandela},
  {Aumont}, {Baccigalupi}, {Banday}, \& et~al.}]{planck2014}
{Planck Collaboration}, {Ade}, P.~A.~R., {Aghanim}, N., {et~al.} 2014, \aap,
  571, A16

\bibitem[{{Riechers}(2013)}]{riechers2013}
{Riechers}, D.~A. 2013, \apjl, 765, L31

\bibitem[{{Riechers} {et~al.}(2009){Riechers}, {Walter}, {Bertoldi}, {Carilli},
  {Aravena}, {Neri}, {Cox}, {Wei{\ss}}, \& {Menten}}]{riechers2009}
{Riechers}, D.~A., {Walter}, F., {Bertoldi}, F., {et~al.} 2009, \apj, 703, 1338

\bibitem[{{Sargsyan} {et~al.}(2014){Sargsyan}, {Samsonyan}, {Lebouteiller},
  {Weedman}, {Barry}, {Bernard-Salas}, {Houck}, \& {Spoon}}]{sargsyan2014}
{Sargsyan}, L., {Samsonyan}, A., {Lebouteiller}, V., {et~al.} 2014, \apj, 790,
  15

\bibitem[{{Schaerer} {et~al.}(2015){Schaerer}, {Boone}, {Zamojski}, {Staguhn},
  {Dessauges-Zavadsky}, {Finkelstein}, \& {Combes}}]{schaerer2015}
{Schaerer}, D., {Boone}, F., {Zamojski}, M., {et~al.} 2015, \aap, 574, A19

\bibitem[{{van der Werf} {et~al.}(2011){van der Werf}, {Berciano Alba},
  {Spaans}, {Loenen}, {Meijerink}, {Riechers}, {Cox}, {Wei{\ss}}, \&
  {Walter}}]{vanderwerf2011}
{van der Werf}, P.~P., {Berciano Alba}, A., {Spaans}, M., {et~al.} 2011, \apjl,
  741, L38

\bibitem[{{Venemans} {et~al.}(2012){Venemans}, {McMahon}, {Walter}, {Decarli},
  {Cox}, {Neri}, {Hewett}, {Mortlock}, {Simpson}, \& {Warren}}]{venemans2012}
{Venemans}, B.~P., {McMahon}, R.~G., {Walter}, F., {et~al.} 2012, \apjl, 751,
  L25

\bibitem[{{Venemans} {et~al.}(2015){Venemans}, {Ba{\~n}ados}, {Decarli},
  {Farina}, {Walter}, {Chambers}, {Fan}, {Rix}, {Schlafly}, {McMahon},
  {Simcoe}, {Stern}, {Burgett}, {Draper}, {Flewelling}, {Hodapp}, {Kaiser},
  {Magnier}, {Metcalfe}, {Morgan}, {Price}, {Tonry}, {Waters}, {AlSayyad},
  {Banerji}, {Chen}, {Gonz{\'a}lez-Solares}, {Greiner}, {Mazzucchelli},
  {McGreer}, {Miller}, {Reed}, \& {Sullivan}}]{venemans2015}
{Venemans}, B.~P., {Ba{\~n}ados}, E., {Decarli}, R., {et~al.} 2015, \apjl, 801,
  L11

\bibitem[{{Walter} {et~al.}(2004){Walter}, {Carilli}, {Bertoldi}, {Menten},
  {Cox}, {Lo}, {Fan}, \& {Strauss}}]{walter2004}
{Walter}, F., {Carilli}, C., {Bertoldi}, F., {et~al.} 2004, \apjl, 615, L17

\bibitem[{{Walter} {et~al.}(2009{\natexlab{a}}){Walter}, {Riechers}, {Cox},
  {Neri}, {Carilli}, {Bertoldi}, {Weiss}, \& {Maiolino}}]{walter2009}
{Walter}, F., {Riechers}, D., {Cox}, P., {et~al.} 2009{\natexlab{a}}, \nat,
  457, 699

\bibitem[{{Walter} {et~al.}(2009{\natexlab{b}}){Walter}, {Wei{\ss}},
  {Riechers}, {Carilli}, {Bertoldi}, {Cox}, \& {Menten}}]{walter2009b}
{Walter}, F., {Wei{\ss}}, A., {Riechers}, D.~A., {et~al.} 2009{\natexlab{b}},
  \apjl, 691, L1

\bibitem[{{Walter} {et~al.}(2003){Walter}, {Bertoldi}, {Carilli}, {Cox}, {Lo},
  {Neri}, {Fan}, {Omont}, {Strauss}, \& {Menten}}]{walter2003}
{Walter}, F., {Bertoldi}, F., {Carilli}, C., {et~al.} 2003, \nat, 424, 406

\bibitem[{{Walter} {et~al.}(2012){Walter}, {Decarli}, {Carilli}, {Riechers},
  {Bertoldi}, {Wei{\ss}}, {Cox}, {Neri}, {Maiolino}, {Ouchi}, {Egami}, \&
  {Nakanishi}}]{walter2012b}
{Walter}, F., {Decarli}, R., {Carilli}, C., {et~al.} 2012, \apj, 752, 93

\bibitem[{{Wang} {et~al.}(2013){Wang}, {Wagg}, {Carilli}, {Walter}, {Lentati},
  {Fan}, {Riechers}, {Bertoldi}, {Narayanan}, {Strauss}, {Cox}, {Omont},
  {Menten}, {Knudsen}, {Neri}, \& {Jiang}}]{wang2013}
{Wang}, R., {Wagg}, J., {Carilli}, C.~L., {et~al.} 2013, \apj, 773, 44

\bibitem[{{Watson} {et~al.}(2015){Watson}, {Christensen}, {Knudsen}, {Richard},
  {Gallazzi}, \& {Micha{\l}owski}}]{watson2015}
{Watson}, D., {Christensen}, L., {Knudsen}, K.~K., {et~al.} 2015, \nat, 519,
  327

\bibitem[{{Willott} {et~al.}(2015){Willott}, {Bergeron}, \&
  {Omont}}]{willott2015}
{Willott}, C.~J., {Bergeron}, J., \& {Omont}, A. 2015, \apj, 801, 123

\bibitem[{{Willott} {et~al.}(2013){Willott}, {Omont}, \&
  {Bergeron}}]{willott2013b}
{Willott}, C.~J., {Omont}, A., \& {Bergeron}, J. 2013, \apj, 770, 13

\bibitem[{{Wu} {et~al.}(2015){Wu}, {Wang}, {Fan}, {Yi}, {Zuo}, {Bian}, {Jiang},
  {McGreer}, {Wang}, {Yang}, {Yang}, {Thompson}, \& {Beletsky}}]{wu2015}
{Wu}, X.-B., {Wang}, F., {Fan}, X., {et~al.} 2015, \nat, 518, 512

\end{thebibliography}
\end{document}